\newcommand{\CLASSINPUTbottomtextmargin}{1.0in}
\newcommand{\CLASSINPUTtoptextmargin}{0.8in}
\documentclass[lettersize,journal]{IEEEtran}
\usepackage{amsmath,amsfonts}
\usepackage{algorithmicx,algorithm}

\usepackage{algorithm,algpseudocode}

\usepackage{array}
\usepackage[caption=false,font=normalsize,labelfont=sf,textfont=sf]{subfig}
\usepackage{textcomp}
\usepackage{amssymb}
\usepackage{stfloats}
\usepackage{url}
\usepackage{verbatim}
\usepackage{graphicx}
\usepackage{cite,soul,xcolor}
\usepackage{bm}      
\newtheorem{theorem}{Theorem}[section]

\newtheorem{lemma}[theorem]{Lemma}
\usepackage{diagbox}
\usepackage{multirow}
\usepackage{booktabs}
\usepackage{tablefootnote}
\usepackage{threeparttable}

\newenvironment{NewProof}{{\noindent\it Proof.}}{\hfill $\blacksquare$\par}

\begin{document}

\title{MIMO Channel as a Neural Function: Implicit Neural Representations for Extreme CSI Compression in Massive MIMO Systems}

\author{Haotian Wu,
Maojun Zhang,
                     Yulin Shao,
                     Krystian Mikolajczyk,
                     Deniz G\"{u}nd\"{u}z
\thanks{H. Wu, Y. Shao, K. Mikolajczyk, and D. G\"und\"uz are with the Department of Electrical and Electronic Engineering, Imperial College London, London SW7 2AZ, U.K. (e-mails: \{haotian.wu17, y.shao, k.mikolajczyk, d.gunduz\}@imperial.ac.uk); M. Zhang is with the College of Information Science and Electronic Engineering, Zhejiang University, Hangzhou 310027, China (e-mail: zhmj@zju.edu.cn). This work is carried out during M. Zhang's visit to Imperial College London.}

\thanks{
}
}



\maketitle

\begin{abstract}
Acquiring and utilizing accurate channel state information (CSI) can significantly improve transmission performance, thereby holding a crucial role in realizing the potential advantages of massive multiple-input multiple-output (MIMO) technology. Current prevailing CSI feedback approaches improve precision by employing advanced deep-learning methods to learn representative CSI features for a subsequent compression process. Diverging from previous works, we treat the CSI compression problem in the context of implicit neural representations. Specifically, each CSI matrix is viewed as a neural function that maps the CSI coordinates (antenna number and subchannel) to the corresponding channel gains. Instead of transmitting the parameters of the implicit neural functions directly, we transmit modulations based on the CSI matrix derived through a meta-learning algorithm. Modulations are then applied to a shared base network to generate the elements of the CSI matrix. Modulations corresponding to the CSI matrix are quantized and entropy-coded to further reduce the communication bandwidth, thus achieving extreme CSI compression ratios. Numerical results show that our proposed approach achieves state-of-the-art performance and showcases flexibility in feedback strategies.
\end{abstract}


\section{Introduction}
Massive multiple-input multiple-output (MIMO) technology stands as the core technology within the landscape of sixth-generation (6G) wireless communication systems and beyond, primarily characterized by its ultra-high reliability, ultra-low latency, and outstanding data rate \cite{9961131}. To achieve the promised data rates, it is essential for the base station (BS) to acquire precise downlink channel state information (CSI). This is achieved by the user equipment (UE) estimating the downlink CSI from the pilots transmitted by the BS, and feeding it back to the BS. However, transmitting the estimated CSI matrices via the feedback link imposes a significant communication overhead, primarily due to the massive number of antennas and sub-carriers available in a typical MIMO system employing orthogonal frequency division multiplexing (OFDM). Hence, it is crucial to develop robust CSI compression techniques to reduce the communication overhead, particularly in scenarios with limited communication bandwidth.

\begin{figure}[t] 
    \centering
    \includegraphics[scale=0.57]{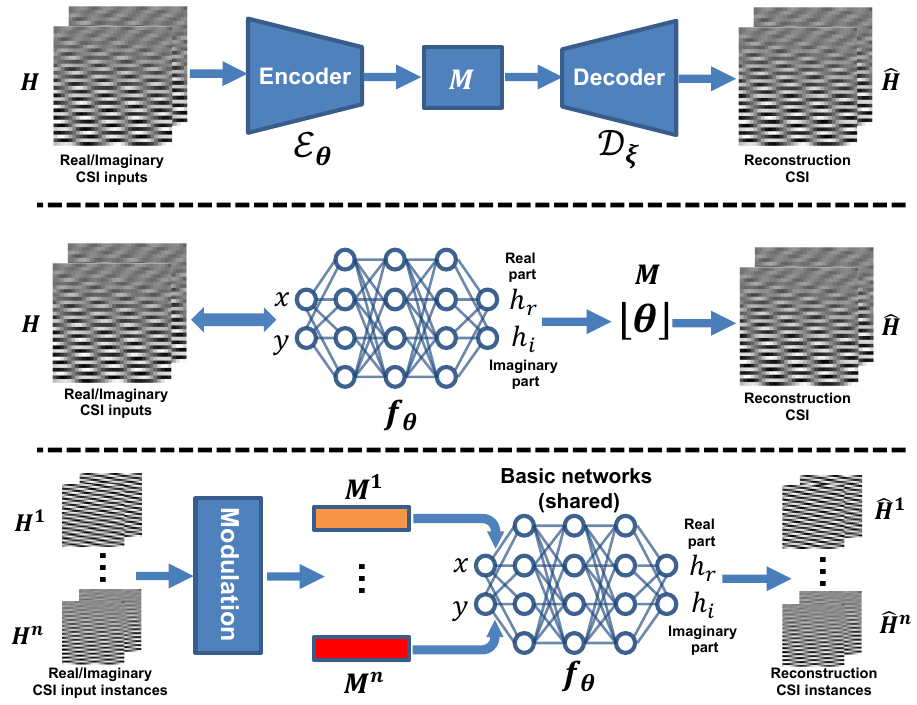}
    \caption{Illustration of alternative CSI feedback methodologies in a MIMO-OFDM system, involving the compression of the CSI matrix $\bm{H}$ into a codeword index $\bm{M}$, which is transmitted to the BS and used to reconstruct the estimated CSI matrix $\bm{\hat{H}}$. Top: Block diagram of feature learning-based methods \cite{guo2022overview,sun2020ancinet,wen2018deep,guo2020convolutional,cao2021lightweight,sun2021lightweight,ji2021clnet,tang2022dilated,cui2022transnet,chen2021deep}, where a pair of encoder and decoder is adopted to learn a low-dimensional latent representation for compression. Middle: Illustration of the potential INR-based CSI compression scheme, where a neural network $f_{\bm{\theta}}$ is employed to map the CSI matrix from corresponding coordinates and thus $\bm{H}$ is converted into a neural function $f_{\bm{\theta}}$ with the compressed feedback codewords $\bm{M}\triangleq\bm{\theta}$. Bottom: Illustration of the proposed CSI-INR scheme, where CSI data points are generated via a shared base network, and the specific information of each CSI instance is modulated into the base network as codeword $\bm{M}$.} 
    \label{Methods_pipeline}
\end{figure}

Considering the implicit correlations inherent in the CSI matrix, researchers have proposed CSI dimension reduction techniques based on vector quantization \cite{love2008overview} and compressed sensing \cite{kuo2012compressive}. Nonetheless, the associated overhead of the former approach exhibits a linear relationship with system dimensions, and the latter relies on the sparsity characteristics of CSI data, both of which limit their applications in a massive MIMO system. Recently, deep learning (DL) techniques for CSI feedback have received significant research interest thanks to their flexibility and adaptivity to the correlations in data \cite{mashhadi2020_tutorial, guo2022overview}. CSINet\cite {wen2018deep} was the first to leverage convolutional neural networks (CNNs) to learn low-dimensional features of CSI. Subsequently, a series of studies \cite{guo2020convolutional, chen2021deep, cao2021lightweight, sun2021lightweight, ji2021clnet, tang2022dilated, cui2022transnet} have been carried out to augment both the performance and efficiency of CSI compression, where different DL technologies, such as recurrent neural networks (RNNs)\cite{li2020spatio}, long short term memory (LSTM)\cite{chen2023viewing}, or transformers \cite{ cui2022transnet} are considered. Unlike the aforementioned methodologies grounded in the angular-delay domain channel, \cite{mashhadi2020distributed, chen2023viewing} proposed deep convolutional and LSTM network-based schemes within the space-frequency domain, where the compression ratios can be further optimized through the quantization and entropy coding technologies. 

As illustrated in Fig. \ref{Methods_pipeline}, all previous studies\cite{guo2022overview, sun2020ancinet,wen2018deep, guo2020convolutional,cao2021lightweight,sun2021lightweight,ji2021clnet,tang2022dilated,chen2021deep,cui2022transnet} view CSI as an image or a sequence. Specifically, they employ diverse DL-based methodologies to transform high-dimensional CSI data into a low-dimensional latent feature vector, followed by quantization, entropy coding operations, and the subsequent reconstruction at the receiver end. Although this auto-encoder based framework demonstrates remarkable performance within the domain of image compression, a notable gap exists in the theoretical interpretation of the optimality pertaining to this methodology when applied to CSI data. Specifically, in contrast to images, the spatial coordinates of the CSI matrix demonstrate significant physical relevance, as they maintain a precise correlation with antenna and subcarrier indices, making channel characteristics highly sensitive to index alterations, such as swapping or translation operations. Furthermore, the neighboring components of the CSI matrix typically exhibit substantial numerical disparities, characterized by noticeable discontinuity along their boundaries. These observations motivate us to develop an interpretable compression scheme that incorporates the intrinsic physical characteristics of CSI data into the compression process.

From a physical standpoint, each element of the CSI matrix characterizes an entanglement of multi-path channel responses. The difference in channel responses observed among distinct antennas and subcarriers primarily arises from the phase difference associated with the propagation delays of individual propagation paths. Considering the antennas' equidistant and compact arrangement, the phase variations between adjacent antennas demonstrate a notable degree of uniformity. Consequently, channel responses can be effectively characterized by a function dependent on the spatial coordinates. Analogous findings can be observed within the frequency domain, resulting from the equidistant subcarrier attributes. Based on the above insights, this paper reconceptualizes CSI as a neural function of the spatial coordinates with an implicit neural representation (INR), which exhibits an impressive capability of over-fitting to the input data distribution. 

Similarly to the idea of per-instance optimization in image compression \cite{van2021overfitting, yang2020improving}, INR-based compression \cite{ronen2019convergence, sitzmann2020implicit} is an emerging technology and can be regarded as an even more extreme instance-specific optimization approach, which optimizes a neural function $f_{\bm{\theta}}$ to overfit every single data point and transmits the resultant network weights $\bm{\theta}$ as the compressed representation of the source data, as illustrated in Fig. \ref{Methods_pipeline}. Recent research \cite{mildenhall2021nerf,sitzmann2020implicit,dupont2022coin,park2019deepsdf,chen2019learning,tancik2020fourier} has shown that using sinusoidal encoding and periodic activation can efficiently model the relationship between spatial locations and the corresponding values of the source signal with high-frequency information, especially for data with complex distributions, such as image or point cloud data \cite{mildenhall2021nerf, sitzmann2020implicit, tancik2020fourier}. 

INR is typically implemented with lightweight architectures of quantized weights, where models are carefully designed with a relatively small number of parameters, resulting in significantly reduced storage when compared to conventional compression approaches. The compression efficiency of this paradigm can be further improved through advanced model compression techniques \cite{havasi2018minimal, van2020bayesian}. More recently, the compression cost of the INR-based approach has been significantly reduced via meta-learning \cite{finn2017model}. Instead of reducing the model size, researchers leverage meta-learning so that only the difference from a shared base network needs to be communicated \cite{dupont2022coin}. Specifically, a shared base network is optimized to generate distinct data based on the input coordinates and different modulations of distinct data points. Meta-learning is adopted to fit the data with the base network within a few steps, where specific information of each instance is then modulated into a base network. 

Recent developments in neural compression, combined with the observed correlations between the INR model and CSI data, motivate the utilization of INR for CSI compression. Specifically, since CSI data can be efficiently expressed as a function of spatial coordinates, INR-compression methodology emerges as a promising approach for CSI representation and communication. In this paper, we propose an INR-based CSI compression approach, dubbed CSI-INR, as shown in Fig. \ref{Methods_pipeline}. 

Our main contributions are summarized as follows:
\begin{itemize}
\item 
Through the analysis and mathematical derivation of the MIMO-OFDM transmission procedures, we reveal that the unique mapping characteristics between the channel gain and its corresponding index naturally align with the INR representation strategy. This finding serves as a theoretical foundation for developing our proposed methodology to view CSI maps as neural functions. 

\item 
Specifically, we propose a novel CSI feedback framework called CSI-INR, where the CSI is conceptualized as a neural function mapping the spatial coordinates to the CSI matrix. To the best of the authors’ knowledge, this is the first INR-based CSI feedback approach and provides a new paradigm for CSI compression in future wireless networks. Notably, this approach provides a physically meaningful representation of CSI, distinct from the prevailing feature learning-based methods that typically view CSI as either an image or sequential data.

\item 
We comprehensively analyze the INR-based CSI feedback framework, exploring different model architectures, modulation techniques, meta-learning strategies, quantization methods, and entropy coding properties. Numerical experiments demonstrate that our proposed model yields significant performance enhancements compared to existing CNN or transformer-based methodologies.

\item 
Moreover, the proposed CSI-INR method exhibits remarkable quantizability and is characterized by its adaptable feedback strategies. Specifically, the CSI-INR framework can support various quantization levels while maintaining minimal performance degradation. This suggests robust performance even under extreme compression ratios. Additionally, users/agents can flexibly select feedback strategies (encoding steps) tailored to their computational resources.
\end{itemize}

\section{Problem Overview}
\subsection{System model}
We consider a massive MIMO system, where a BS, equipped with $N_t\gg 1$ antennas in a uniform linear array (ULA) configuration serves a single-antenna UE using OFDM modulation over $N_c$ subcarriers. The downlink received signal at the UE, $\bm{y}\in \mathbb{C}^{N_c}$, can be written as \cite{9718553,mashhadi2020distributed}:
\begin{equation}
    \bm{y}=\bm{H}^H\bm{v}x+\bm{z},
\end{equation}
where $\bm{H}\in \mathbb{C}^{N_t\times N_c}$ is the channel matrix, $\bm{v}\in \mathbb{C}^{N_t}$ is the precoding vector, ${x}\in \mathbb{C}$ is the data-bearing symbol, and $\bm{z}\in \mathbb{C}^{N_c}$ is the additive white Gaussian noise vector. The downlink channel matrix $\bm{H}$ is estimated at the UE side and fed back to the BS.

Let us write
\begin{equation}
   \bm{H}\triangleq\left[\bm{h}(f_1), \bm{h}(f_2), \ldots, \bm{h}(f_{N_c})\right], 
\end{equation}
where $\bm{h}(f_i)\in \mathbb{C}^{N_t}$, $i=1,\dots,N_c$, is the channel gain vector of the $i$-th subcarrier with frequency $f_i$. We define $f_i=f_0+(i-1)f_{\triangle}$, where $f_0$ is the lowest subcarrier frequency, and $f_{\triangle}$ is the subcarrier spacing. In particular, the channel gain vector $\bm{h}(f)$ can be modeled as \cite{8795533}:
\begin{equation}
    \bm{h}(f)=\sum_{p=1}^{P}\alpha_{p}e^{-j2\pi f\tau_p+j\phi_p}\bm{a}(\theta_p),
    \label{eq_h}
\end{equation}
where $P$ is the total number of paths, $f$ is the carrier frequency, $\alpha_p$, $\tau_p$, $\phi_p$ and $\theta_p$ are the propagation gain, propagation delay, initial phase shift, and angle of departure (AoD) of the $p$-th path, respectively. Moreover, $\bm{a}(\theta_p)\in \mathbb{C}^{N_t}$ is the steering vector given by
\begin{equation}
    \bm{\alpha}(\theta_p)=\left[1,e^{-j\chi\sin\theta_p}, \ldots, e^{-j\chi(N_t-1)\sin\theta_p}\right]^T,
        \label{eq_a}
\end{equation}
where $\chi\triangleq\frac{2\pi df}{c}$, $d$ is the antenna spacing, and $c$ is the speed of light. 

\subsection{DL-based CSI compression}
Considering the constraints imposed by the limited bandwidth of the feedback link and the high transmission cost of $\bm{H}$ in a massive MIMO-OFDM system, state-of-the-art schemes employ learning-based techniques to compress $\bm{H}$ into a low dimensional representation $\bm{M}\in\mathbb{R}^n$. The codeword index is transmitted to the receiver, from which the BS reconstructs $\bm{\hat{H}} \in \mathbb{C}^{N_t\times N_c}$. For such a system, we define the compression ratio as $$CR\triangleq\frac{n}{2N_tN_c}.$$

Prior CSI feedback schemes\cite{guo2022overview,sun2020ancinet,wen2018deep,guo2020convolutional,cao2021lightweight,sun2021lightweight,ji2021clnet,tang2022dilated,chen2021deep,li2020spatio,chen2023viewing,cui2022transnet,mashhadi2020distributed} primarily employ feature learning-based methodologies. The essence of such schemes is to treat the CSI compression scheme as an autoencoder with an encoder at the UE and a decoder at the BS.
The encoder, denoted by $\mathcal{E}(\cdot): \mathbb{C}^{N_t\times N_c} \rightarrow \mathbb{R}^{n}$, maps the CSI matrix directly into the low-dimensional latent features $\bm{M}\in \mathbb{R}^{n}$.
The decoder, denoted by $\mathcal{D}(\cdot): \mathbb{R}^n \rightarrow\mathbb{C}^{N_t\times N_c}$ aims to reconstruct the CSI matrix with minimal distortion from the received latent representation. Both the encoder and the decoder are parameterized by deep neural networks (DNN), yielding
\begin{equation}
\begin{array}{cc}
    \bm{M}=\mathcal{E}_{\bm{\theta}}(\bm{H}), & \bm{\hat{H}}=\mathcal{D}_{\bm{\xi}}(\bm{M}),
\end{array}
\end{equation}
where $\bm{\theta}$ and $\bm{\xi}$ are the DNN parameters of $\mathcal{E}(\cdot)$ and $\mathcal{D}(\cdot)$, respectively.
Given a distortion measure $\mathcal{L}(\bm{H},\bm{\hat{H}};\bm{\theta},\bm{\xi})$, the encoder and decoder are jointly trained through gradient descent by treating the distortion as the loss function.

For the feature learning-based approaches, the reconstruction performance is mainly determined by the architecture of the DNNs and the available bandwidth. Diverse DNN architectures, such as CNNs \cite{guo2020convolutional}, transformers \cite{9961131, cui2022transnet}, and LSTM networks \cite{chen2023viewing} have been utilized in the literature. The latent feature vector $M$ encapsulates the entire semantics and information of the CSI data, the length of which reflects the necessary bandwidth and is of vital importance for reconstruction performance. It is worth noting that, while feature extraction serves for dimensionality reduction, the compression framework requires mapping the features to bits to be transmitted over the feedback link. While vector $\bm{M}$ can be communicated effectively using a high-resolution representation (e.g., 32-bit floating point) for each of its elements, a more efficient compression scheme can be obtained by employing quantization and entropy coding techniques to further reduce the bandwidth requirements.

\subsection{INR-based CSI compression}
Drawing on the concept of INR \cite{strumpler2022implicit,gordon2023quantizing,guo2023compression}, this paper puts forth a novel CSI feedback scheme dubbed CSI-INR. In contrast to the prevailing feature learning-based methodologies, the cornerstone of our INR-based approach lies in representing the CSI matrix by a neural function $f_{\bm{\theta}}: \mathbb{R}^{2} \rightarrow \mathbb{C}$, which establishes a mapping from the coordinates $(i,j)$ to the corresponding channel gains $H[i,j]$, with $\bm{\theta}$ denoting the parameters of the neural function. For any given instance of the random CSI matrix $\bm{H}$, represented as $\bm{H}^i$,\footnote{In this paper, the notation $\bm{X}^i$ with an upper index $i=1,2, \ldots$ indicates a specific instance of the random matrix $\bm{X}$.} a unique neural function with parameters $\bm{\theta}^i$ can be derived. In so doing, our emphasis transitions from transmitting the feature vector directly extracted from the CSI matrix to transmitting the neural function, specifically the parameters $\bm{\theta}^i$. Upon receiving $\bm{\theta}^i$, the BS can reconstruct the channel gains of the CSI matrix by iterating over its coordinates. By keeping the number of parameters $|\bm{\theta}|$ to a minimum, our new approach has the potential to deliver substantially improved rate-distortion performance.



The comprehensive architecture of our CSI-INR approach is illustrated in Fig.~\ref{Methods_pipeline}. As opposed to identifying a unique neural function for each CSI matrix instance $\bm{H}$, we decompose the neural function into two parts: a base neural function and a CSI modulation function.
The base neural function, denoted by $f_{\bm{\theta}^*}$, is shared between the UE and the BS, with $\bm{\theta}^*$ representing a set of meta-learned initial parameters. Its primary role is reconstruction, i.e., mapping the coordinate matrix $\bm{x}\in \mathbb{R}^ {N_t\times N_c\times 2}$, whose element $\bm{x}[i,j]$ is the coordinate pair $(i,j)$, to the channel gain matrix $\bm{H}$.
On the other hand, the CSI modulation function, denoted by $\mathcal{M}(\cdot)$, is deployed only at the UE side.
For an instance of the CSI matrix $\bm{H}^i$, the CSI modulation function transforms it into a vector $\bm{M} \in \mathbb{R}^{n}$, which is then transmitted to the BS. Upon receipt, the BS modifies certain parameters in the shared base neural function $f_{\bm{\theta}^*}$ with the codeword $\bm{M}$, tailoring it specifically to the CSI instance $\bm{H}^i$. The above operations can be encapsulated as follows:
\begin{equation}
\begin{array}{cc}
    \bm{M}=\mathcal{M}(\bm{H}), & \bm{\hat{H}}={f}_{\bm{\theta}^*}(\bm{x},\bm{M}).
\end{array}
\label{sys_formulation}
\end{equation}



\section{CSI with Implicit Neural Representations}
This section starts with a theoretical elucidation of representing the CSI data with implicit neural networks. 
Subsequently, we present the architecture of the proposed CSI-INR scheme. Following this, we explore the incorporation of quantization and entropy coding operations within the CSI-INR scheme to further enhance its performance. Finally, we provide a summary of the proposed CSI feedback strategy.

\subsection{Mapping the channel gains from spatial coordinates}

From a theoretical perspective, each element within the CSI matrix arises from the coupling of multi-path channel responses. Variations in channel responses across different antennas and subcarriers primarily stem from phase alterations associated with the propagation delays of individual propagation paths. These phase distinctions can also be, in turn, indicative of spatial coordinates of elements, and hence, can be formulated as a function over the spatial coordinates of the channel gain matrix. Specifically, we reveal the detailed mapping relationship between the channel gain and corresponding coordinates in the following lemma:

\begin{lemma}
There exists a function that can physically map each element of the channel matrix ${H}[n,m]$ from the corresponding coordinate pair $(n,m)$, represented as ${H}[n,m]=f_{\bm{H}}(n,m)$. Consequently, the channel matrix can be comprehensively characterized as a result of the mapping process applied to a series of coordinate pairs as:
\begin{equation}
\bm{H}=\left[
\begin{array}{cccc}
f_{\bm{H}}(1,1)&f_{\bm{H}}(1,2)&\cdots&f_{\bm{H}}(1,N_c)\\
f_{\bm{H}}(2,1)&f_{\bm{H}}(2,2)&\cdots&f_{\bm{H}}(2,N_c)\\
\vdots&\vdots&\ddots&\vdots\\
f_{\bm{H}}(N_t,1)&f_{\bm{H}}(N_t,2)&\cdots&f_{\bm{H}}(N_t,N_c)\\
\end{array}
\right].
\label{lemma1_eq}
\end{equation}
\label{lemma1}
\end{lemma}

\begin{NewProof}
Based on Eqn. \eqref{eq_h} and Eqn. \eqref{eq_a}, we can get the $n$-th row and $m$-th column of $\bm{H}$ as:
\begin{align}
    {H}[n,m]=\sum_{p=1}^{P}\alpha_{p}e^{-j2\pi (f_0+(m-1)f_{\triangle}) \tau_p+j\phi_p}e^{-j\chi(n-1)\sin\theta_p}\notag\\
    =\sum_{p=1}^{P}\alpha_{p}e^{-j2\pi f_0 \tau_p+j\phi_p}e^{-j[2\pi(m-1)f_{\triangle}\tau_p+\chi(n-1)\sin\theta_p]}.
\end{align}    
Considering the parameters of the $p$-th path $\alpha_p$, $f_0$, $\tau_p$, $\phi_p$, $\theta_p$ and $\mathcal{X}$ are determined by the transmission system setting, the coefficient factor of the $p$-th path $A_p\triangleq \alpha_{p}e^{-j2\pi f_0 \tau_p+j\phi_p}$ is independent of the coordinate $(n,m)$. Thus, we can write each element of $\bm{H}$ as a function of its coordinates $(n,m)$ as: 
\begin{align}
    {H}[n,m]&=\sum_{p=1}^{P}A_{p}e^{-j[2\pi(m-1)f_{\triangle}\tau_p+\chi(n-1)\sin\theta_p]}\notag\\
    &=f_{\bm{H}}(n,m),
\end{align}
yielding Eqn. \ref{lemma1_eq}.
\end{NewProof}

\begin{figure*}[t] 
    \centering
    \includegraphics[scale=0.688]{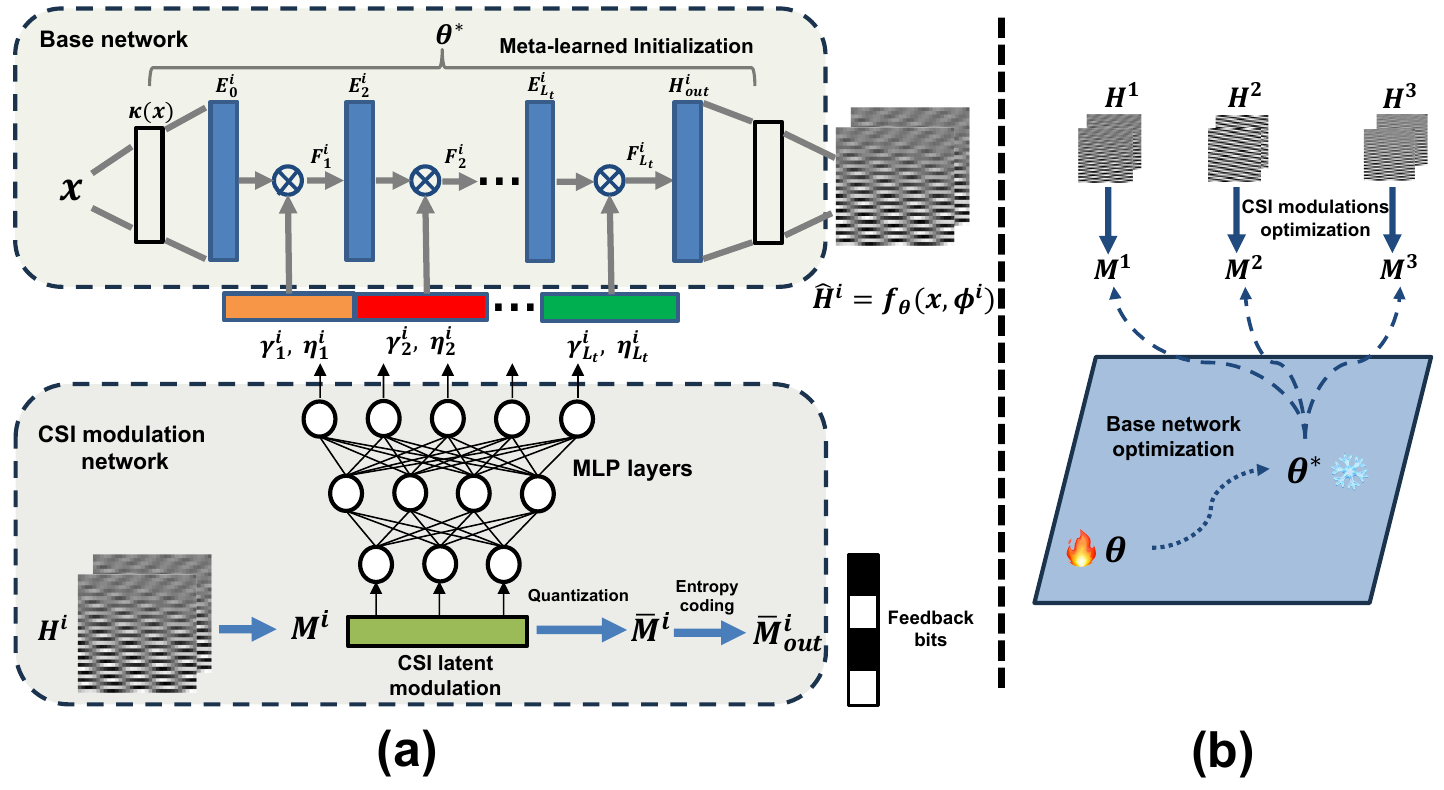}
    \caption{(a). Basic blocks of our CSI-INR structure, where each data point $\bm{H^i}$ is modulated into $\bm{M^i}$ to generate the modulation factors $\bm{\phi^i}\triangleq\left[ \bm{\phi^i_1},\cdots,\bm{\phi^i_{L_{t}}}\right]$, each element of which is defined as $\bm{\phi^i_k}\triangleq\left[\bm{\gamma^i_k}, {\bm{\eta^i_k}}\right]$, representing the modulation factors of $i$-th sample for the $k$-th layer of the base network. The FiLM modulation operations are employed with the above factors and denoted as $\bigotimes$. Modulations are operated over the intermediate feature output $\bm{E^i_k}$ of the base network, resulting in modulated features $\bm{F^i_k}$, to map the coordinate input $\bm{x}$ into the distinct CSI data $\bm{H^i}$. (b). The illustration of the training process of the CSI-INR scheme. Specifically, the network $\bm{\theta}$ is optimized through a meta-learning strategy, where $\bm{\theta}$ and $\bm{M}$ are frozen and optimized in turns until it converges into the optimal network parameters $\bm{\theta^*}$.}
    \label{architecture}
\end{figure*}

Lemma \ref{lemma1} reveals the physical interpretation of a mapping from the spatial coordinates to the corresponding elements of the channel matrix. Specifically, we can observe that the difference between adjacent antennas or subcarriers is a fixed phase term $\mathcal{X}\sin\theta{_p}$ and $2\pi f_{\triangle}\tau_p$ for the $p$-th path. The complete matrix can then be conceptualized as a complex-valued proportional mapping across diverse pathways for individual rows and columns. 

In light of the characteristics of the inherent correlation among elements within the CSI matrix, we propose an INR-based model to capture and model this implicit relationship. Specifically, we adopt neural functions $f_{\bm{\theta}}(\cdot)$ with parameters $\bm{\theta}$ to fit the mapping function $f_{\bm{H}}(\cdot)$, where  $f_{\bm{\theta}}(\cdot)$ is optimized to map the coordinates to the corresponding channel gain elements.
 
\subsection{The architecture of the CSI-INR scheme}
As shown in  Fig. \ref{architecture} (a), the CSI-INR scheme predominantly comprises a base network, parameterized by ${\bm{\theta}}$, alongside a CSI modulation network. This section will expound upon the CSI-INR model with the following principal components: the base INR network, CSI modulation, and training strategy.

\subsubsection{Base INR network}
The base INR network is a shared network architecture across all data points, parameterized by $f_{\bm{\theta}}(\cdot): \mathbb{R}^2\rightarrow\mathbb{C}$. This network functionally maps a set of spatial coordinates $\bm{x}=\left[n,m\right]^T$ within the CSI matrix $\bm{H}$ to their associated channel gains $H[n,m]$. Inspired by the representational power of the Fourier Feature Network (FFN) \cite{tancik2020fourfeat} and the fitting capability of sinusoidal representation networks (SIREN) \cite{chu1972polyphase,sitzmann2020implicit} for INR, our base network is designed with a Fourier Transform block and $L_t$ SIREN layers, as illustrated in Fig. \ref{detailed_architecture}. 

To be specific, the Fourier Transform block, denoted by $\kappa(\bm{x}): \mathbb{R}^2\rightarrow\mathbb{R}^{2d}$, can enhance the representation and mitigate the spectral bias in coordinate-based multilayer perceptron (MLP) networks when modeling high-frequency functions in lower dimensions \cite{tancik2020fourfeat}. The operations of FFN can be expressed as:
\begin{equation}
\kappa(\bm{x})=\left[
\begin{array}{c}
  \cos(2\pi\bm{B}\bm{x})\\
\sin(2\pi\bm{B}\bm{x}) 
\end{array}
\right],
\end{equation}
where $\bm{x}\in \mathbb{R}^{2}$ is the input coordinate vector, $\bm{B}\in \mathbb{R}^{d\times 2}$ is the frequency vector, with each of its components drawn from a Gaussian distribution $\mathcal{N}(0,\sigma_b^2)$, and $d$ is the dimension of the intermediate feature map. 

Meanwhile, to accurately model the signals with fine details and to capture both spatial and temporal derivatives, we incorporate $L_t$ SIREN layers subsequent to the FFN block, where the $i$-th SIREN layer, $i=1,\cdots,L_t$, is represented by:
\begin{equation}
    \bm{E_i}=\sin(\omega_0(\bm{W_i}\bm{E_{i-1}}+\bm{b_i})),
    \label{SIREN}
\end{equation}
where $\bm{W_1}\in \mathbb{R}^{d\times 2d}$ and $\bm{W_i}\in \mathbb{R}^{d\times d}$, for $i=2,\cdots,L_t$, are weights of the linear layer, $\bm{b_i}\in \mathbb{R}^{d}$ is the bias vector, and the sinusoidal function with parameter $\omega_0$ is the activation function. $\bm{E_i}$ is the output feature map of $i$-th SIREN layer, where $\bm{E_0}\triangleq\kappa(\bm{x})\in \mathbb{R}^{2d}$ and $\bm{E_i}\in \mathbb{R}^{d}$ for $i=2,\cdots, L_t$.

From a conventional INR perspective, the above base network can be optimized to facilitate the reconstruction of individual channel gains based on their corresponding coordinates with the help of a linear layer, which can map the feature matrix $\bm{E_{L_t}}$ into the channel gain matrix $\bm{\hat{H}}$. This involves the fine-tuning of distinct network parameters to align with the unique characteristics of each data point. As a result, this procedure generates a collection of customized network configurations corresponding to various data points. The codeword of $i$-th data point is subsequently defined as the parameters of the corresponding network, denoted by $\bm{M^i}\triangleq\bm{\theta^i}$.

\subsubsection{CSI modulation}
Rather than fine-tuning different neural networks for distinct CSI data, we implement a modulation strategy on each individual CSI matrix. In particular, we transform each data point into a modulation vector to adjust the base network, which induces variations to the parameters of the common base network shared across all data points. This approach facilitates the parameterization of data points into families of neural networks.

To be more specific, we employ the feature-wise linear modulation (FiLM) strategy \cite{perez2018film} as:
\begin{equation}
    \text{FiLM}(\bm{E})=\bm{\gamma}\cdot\bm{E}+\bm{\eta},
\end{equation}
where $\bm{\gamma}\in\mathbb{R}^d$ and $\bm{\eta}\in\mathbb{R}^d$ are the elementwise scale factors and shifts derived from a modulation network, and $\bm{E}\in\mathbb{R}^d$ is the intermediate feature from the base network. Different from the modulation methods in \cite{chan2021pi, mehta2021modulated}, we adopt the FiLM modulation strategy on the intermediate output of each SIREN layer within the base network, yielding a modulated SIREN block, as shown in Fig. \ref{detailed_architecture}. In particular, each modulated SIREN block compromises a SIREN layer and a modulation block, where the modulation block modulates the result of the corresponding SIREN block via FiLM strategy. To note, different modulation vectors derived from MLP layers are applied to different SIREN layers. For each data point, the output of the $i$-th modulated SIREN layer $\bm{F_i}$ can be represented as follows:
 \begin{figure}[t] 
    \centering
    \includegraphics[scale=0.31]{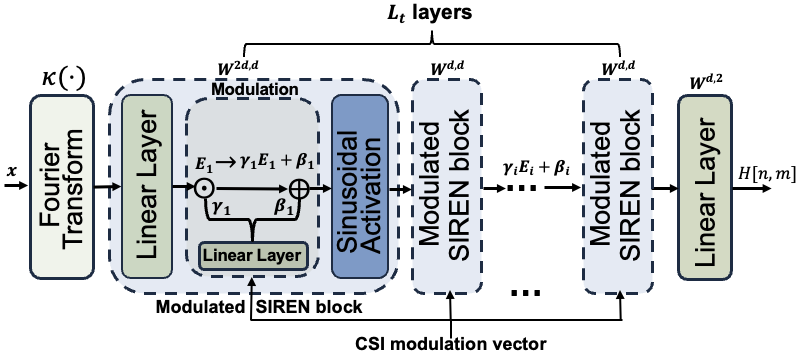}
    \caption{Inner architecture of the CSI-INR model, where $\bigodot$ is the element-wise multiplication operation and $\bigoplus$ is the adding operation.}
    \label{detailed_architecture}
\end{figure}
\begin{align}
    \bm{F_{i}}
    =&\sin(\omega_0(\bm{\gamma_i}\bm{E_i}+\bm{\eta_i}))\\
    =&\sin(\omega_0(\bm{\gamma_i}(\bm{W_i}\bm{F_{i-1}}+\bm{b_i})+\bm{\eta_i})),
\end{align}        
where $\bm{E_i}\triangleq\bm{W_i}\bm{F_{i-1}}+\bm{b_i}$ is the intermediate output before the modulation given the output of the $(i-1)$-th modulated SIREN layer $\bm{F_{i-1}}$. 

For each data point, the modulated vectors at the $i$-th modulated SIREN layer can be represented as $\bm{\phi_i}\triangleq\left[\bm{\gamma_i}, {\bm{\eta_i}}\right]$, including the modulation scales $\bm{\gamma_i}$ and modulation shifts $\bm{\eta_i}$, derived from a separate modulation network. Specifically, we employ a simple MLP-based network as the modulation network to generate each $\bm{\phi_i}$ for $i=1,\cdots,L_{t}$. Taking one layer as example, the modulation network can be represented as follows:
\begin{equation}
\begin{array}{l}
\bm{\phi_i}\triangleq\left[\bm{\gamma_i}, {\bm{\eta_i}}\right],\\
    \bm{\gamma_i}=\bm{W_{\gamma_{i}}}\bm{M^k}+\bm{b_{\gamma_{i}}},\\
    \bm{\eta_i}=\bm{W_{\eta_{i}}}\bm{M^k}+\bm{b_{\eta_{i}}},
\end{array}
\label{modulate}
\end{equation}
where $\bm{M^{k}}\in \mathbb{R}^{n}$ is the modulated codeword of the $k$-th data point; $\bm{W_{\gamma_{i}}}\in \mathbb{R}^{d\times n}$, $\bm{b_{\gamma_{i}}}\in \mathbb{R}^{d}$, $\bm{W_{\eta_{i}}}\in \mathbb{R}^{d\times n}$, $\bm{b_{\eta_{i}}}\in \mathbb{R}^{d}$ are the linear weights and bias vectors to generate $\bm{\gamma_i}$ and $\bm{\eta_i}$ for the $i$-th modulated SIREN block. To note, $\bm{W_{\gamma_{i}}}$, $\bm{b_{\gamma_{i}}}$, $\bm{W_{\eta_{i}}}$, $\bm{b_{\eta_{i}}}$ are optimized to fit different batches of data points across the whole training set.

After $L_t$ modulated SIREN blocks, a linear layer is connected to map $\bm{F_{L_t}}$ to the final channel gain $\bm{H_{out}}\in\mathbb{R}^2$ of each coordinate as:
\begin{equation}
\bm{H_{out}}=\bm{W_{out}}\bm{F_{L_t}}+\bm{b_{out}}\in\mathbb{R}^2,
\end{equation}
where $\bm{W_{out}}\in\mathbb{R}^{2,d}$, $\bm{b_{out}}\in\mathbb{R}^2$ are the weight and bias of the final linear layer, and the output $\bm{H_{out}}$ is then converted into the corresponding channel gain $H[n,m]\in\mathbb{C}$.

Through incorporating diverse scales and shifts across multiple intermediate SIREN layers, our modulated SIREN layers, combined with an initial Fourier Transform, enable the parameterization of various CSI data points within an ensemble of neural networks. This ensemble demonstrates noteworthy capabilities in compressing and reconstructing CSI data with commendable performance, as evidenced by the results of our numerical experiments presented in Section IV.

\begin{algorithm}[t] 
    \caption{Training strategy of the CSI-INR scheme}
    \textbf{Training phase:} \\
    \textbf{Input:}\\
    Number of data points $N\in \mathbb{R}$,\\
    The $i$-th input sample pair: $\bm{d^i}\triangleq\left\{ (\bm{x},\bm{H^i})\right\}$,\\
    The $i$-th channel matrix sample : $\bm{H^i}\in \mathbb{C}^{N_t\times N_c}$, \\
    Inner optimization step: $s \in \mathbb{R}$ for the training phase,\\
    Inner loop learning rate: $\alpha\in \mathbb{R}$.\\
    Outer loop learning rate: $\beta\in \mathbb{R}$.\\
    \textbf{Output:}\\
    Parameters of the base neural function: $\bm{\theta}$;\\
    Modulated codeword of each data point: $\bm{M_s^i}\in \mathbb{R}^{n}$.\\

    \vspace{-15pt}
    \begin{algorithmic}[1]
    \For{each data point within the batch}
    \State \texttt{$\bm{M_0^i}=\bm{0_{n}}\in \mathbb{R}^{n}$}
    \Comment{Initialize the codeword}
    \For{j= 1:1:$s$}
    \State \texttt{$\bm{M}^i_j=\bm{M}^i_{j-1}-\alpha\nabla_{\bm{M^i_{j-1}}}\mathcal{L}_{mse}(\bm{\theta},\bm{M^i_{j-1}},\bm{d^i})$}\\
    \Comment{Optimize the codeword for each data point}
     \EndFor
     
        \State \texttt{$\bm{\theta}=\bm{\theta}-\beta\nabla_{\theta}\sum_{i=1}^{N}\mathcal{L}_{mse}(\bm{\theta},\bm{M^{i}_{s}},\bm{d^i})$}\\
        \Comment{Optimize the base network for the batch}
                 \EndFor     
                 
    \end{algorithmic}
    \vspace{+3pt}
    \label{alg_train}
\end{algorithm}

\subsubsection{Training strategy}
In order to optimize the base network and encode the CSI matrix into low dimensional modulated representation, we minimize a mean square error (MSE) loss $\mathcal{L}_{mse}$ between the input and its reconstruction:
\begin{align}
&\mathcal{L}_{mse}(\bm{\theta},\bm{M},\bm{d})\notag\\
   &= \frac{1}{NN_tN_c}\sum_{i=1}^N\sum_{n=1}^{N_t}\sum_{m=1}^{N_c}\left \| f_{\bm{\theta}}(\bm{x}(n,m),\bm{M^i})-\bm{H^i}[n,m] \right \|_2^2	
\end{align}
where $\bm{x}[n,m]\triangleq \left[n,m\right]^T$ is the coordinate pair vector, and $\bm{d}\triangleq\left\{ (\bm{x},\bm{H^i})\right\}_{i=1}^N$ is a set comprising $N$ sample pairs of CSI matrices and their corresponding coordinate vectors. Each element $\bm{d}^i \triangleq (\bm{x}, \bm{H}^i)$ within this set refers to the $i$-th instance pair, where $\bm{M}^i$ is the modulated codeword associated with the coordinate $\bm{x}$ and the channel matrix $\bm{H}^i$.

Instead of directly optimizing $\mathcal{L}_{mse}$, we employ the model-agnostic meta-learning (MAML)\cite{finn2017model} method to fit the data with a few gradient steps. Similar approaches are also applied in \cite{sitzmann2020implicit,tancik2020fourfeat,dupont2022coin}. We search for a good initialization $\bm{\theta^*}$ such that the fitting process of minimizing $\mathcal{L}_{mse}$ for each input can be completed in a few gradient steps. The detailed training strategy is summarized in Algorithm \ref{alg_train}, which encompasses an inner training loop (CSI modulation optimization) and an outer training loop (base network optimization), which are iteratively optimized for enhanced performance.
\begin{algorithm}[t] 
    \caption{CSI encoding strategy of the CSI-INR scheme}
    \textbf{Encoding phase:}\\
        \textbf{Input:}\\
     The $i$-th CSI sample pair: $\bm{d^i}\triangleq\left\{ (\bm{x},\bm{H^i})\right\}$,\\
    Base neural function: $f_{\bm{\theta}}(\cdot)$, inner learning rate $\alpha$,\\
    Quantizer: $\mathcal{Q}_q(\cdot)$, entropy encoder $\mathcal{E}_e(\cdot)$,\\
    The inner optimization step number for the inference: $s_{in} \in \mathbb{R}$.\\
    \textbf{Output:}\\
        Modulated codeword: $\bm{M_{s_{in}}^i}\in \mathbb{R}^{n}$,\\
        Modulated bits: $\bm{\bar{M}}_{out}^i\in \mathbb{Z}^{b}$.\\
    \vspace{-15pt}
    \begin{algorithmic}[1]
        \State \texttt{$\bm{M_0^i}=\bm{0_{n}}\in \mathbb{R}^{n}$}
    \Comment{Initialize the codeword}
    \For{j= 0:1:$s_{in}$}
    \State \texttt{$\bm{M}^i_j=\bm{M}^i_{j-1}-\alpha\nabla_{\bm{M^i_{j-1}}}\mathcal{L}_{mse}(\bm{\theta^*},\bm{M^i_{j-1}},\bm{d^i})$}
     \EndFor
     \Comment{Modulate the codeword}
     \If{Quantization and entropy coding are applicable}
     \State \texttt{
     $\bm{\bar{M}}^i=\mathcal{Q}_q(\bm{M_{s_{in}}^i})$}
     \Comment{Quantize the codeword}
     \State \texttt{
     $\bm{\bar{M}}_{out}^i=\mathcal{E}_e(\bm{\bar{M}}^i)$}
     \Comment{Entropy encoding}
      \EndIf
    \end{algorithmic}
    \label{alg_feedback}
\end{algorithm}

In the inner training loop, the primary objective is to optimize a batch of $N$ modulated codewords $\left\{\bm{M^1},\bm{M^2},\cdots,\bm{M^N} \right\}$ via minimizing $\mathcal{L}_{mse}$, while keeping the base network parameters $\bm{\theta}$ fixed. In contrast, the outer training loop focuses exclusively on optimizing $\bm{\theta}$ while keeping the optimized modulated codeword fixed. Specifically, given the optimized modulated codeword batch, the outer training loop is performed to update the base network parameters to minimize $\mathcal{L}_{mse}$.

Starting from an initialization of $\bm{\theta}$ and $\bm{M}$, we first perform $s$ steps of inner loop optimization. A single inner loop optimization step, taking $j$-th step as an example, over each input $\bm{d}^i\triangleq\left\{\bm{x},\bm{H^i}\right\}$ can be given as:
\begin{equation}
    \bm{M_j^i}=\bm{M^i_{j-1}}-\alpha\nabla_{\bm{M^i_{j-1}}}\mathcal{L}_{mse}(\bm{\theta},\bm{M_{j-1}^i},\bm{d^i}),
    \label{inner_loop}
\end{equation}
where $\alpha$ is the learning rate of the inner loop optimization, $\bm{\theta}$ is fixed, and $\bm{M^i_j}$ represents the optimized modulated codeword of $i$-th data sample after the $j$-th inner optimization step. 

Given a batch of $N$ inputs, after executing $s$ steps inner loop optimization as defined by Eqn. \eqref{inner_loop}, the resulting $\bm{M_s^i}$ is obtained for each individual input. Subsequently, an outer loop optimization is conducted according to the following expression:
\begin{equation}
  \bm{\theta}=\bm{\theta}-\beta\nabla_{\bm{\theta}}\sum_{i=1}^{N}\mathcal{L}_{mse}(\bm{\theta},\bm{M^{i}_{s}},\bm{d^{i}}),
\end{equation}
where $\bm{M_s^i}$ is fixed for each input, and $\beta$ is the outer loop learning rate. 

Upon the completion of $\bm{\theta}$ updates in the outer loop, the inner loop of $s$ steps is iteratively performed to update a new modulated codeword $\bm{M_s}$. This process continues iteratively until convergence. This training strategy facilitates the meta-learning procedure of the base network, enabling rapid adaptation for each input. As a result, a well-trained base network has the capability to efficiently optimize the modulated codeword of individual inputs with minimal iterations, achieving competitive performance.
\begin{algorithm}[t] 
    \caption{CSI decoding strategy of the CSI-INR scheme}
    \textbf{Decoding phase} \\
        \textbf{Input:}\\     
        Modulated codeword: $\bm{M_{s_{in}}^i}\in \mathbb{R}^{n}$,\\ Modulated bits $\bm{\bar{M}}_{out}^i\in \mathbb{Z}^{b}$,\\
        Base neural function: $f_{\bm{\theta}}(\cdot)$,\\
    Dequantizer: $\mathcal{D}_{q}(\cdot)$, entropy decoder $\mathcal{D}_e(\cdot)$,\\
        Coordinate vector $\bm{x}$.\\
    \textbf{Output:}\\
    Reconstructed CSI data: ${\bm{\hat{H}}}$\\
    \vspace{-15pt}
    \begin{algorithmic}[1]
      \If{Quantization and entropy coding are applicable}
     \State \texttt{
     $\bm{\bar{M}}^i=\mathcal{D}_e(\bm{\bar{M}}_{out}^i)$}
     \Comment{Entropy decoding}
     \State \texttt{
     $\bm{\hat{M}^i}=\mathcal{D}_q(\bm{\bar{M}^i})$}
     \Comment{Dequantize the codeword}
     \State \texttt{$\bm{\hat{H}}=f_{\bm{\theta}}(\bm{\hat{M}^i},\bm{x})$}
    \ElsIf{}
          \Comment{Direct feedback modulated codeword}
    \State \texttt{$\bm{\hat{H}}=f_{\bm{\theta}}(\bm{M_{s_{in}}^i},\bm{x})$}
     \EndIf
      \Comment{Generate each channel gain}
    \end{algorithmic}
    \label{alg_de}
\end{algorithm}

\subsection{Quantization and entropy coding}
To further enhance the compression performance, we quantize the obtained latent representation and apply entropy coding. Particularly, within the context of a feature learning-based approach, the integration of quantization followed by entropy coding can significantly improve the rate-distortion trade-off as demonstrated in \cite{mashhadi2020distributed}. Concurrently, techniques based on INR can leverage quantization techniques over neural network parameters, as evidenced by \cite{dupont2021generative}. Notably, neural networks can undergo quantization down to $16$ bits without experiencing a substantial decline in performance.

In contrast to previous methodologies, our MAML-based CSI-INR scheme exhibits a notable feature: the modulation codewords associated with each CSI data are surprisingly quantizable. Employing a simple uniform quantization approach, we can compress each element of the codeword for the $i$-th sample after $s$ inner optimization steps into a reduced bitwidth of $b$ bits. The quantization and dequantization process can be represented as:
$\bm{\bar{M}}^i=\mathcal{Q}_q(\bm{M_{s}^i}),\quad\bm{\hat{M}}^i=\mathcal{D}_q(\bm{\bar{M}}^i)$,
where the $\mathcal{Q}_q(\cdot)$ rounds each element of $\bm{M_{s_{in}}^i}$ into the nearest bin, and $\mathcal{D}_q(\cdot)$ dequantizes to the values at the center of each quantization bin, resulting in $\bm{\hat{M}}^i$. Remarkably, our investigation reveals that reducing the bit count from $32$ bits to $3-8$ bits for each element of codewords yields only limited reduction in performance, which will be detailed later in Fig. \ref{exp_quant}. We want to emphasize that the employed uniform quantization approach is a simple quantization method characterized by low computational complexity, and there exists further space to optimize this process for better performance in this further. 

Moreover, entropy coding, such as arithmetic coding \cite{rissanen1979arithmetic}, can further exploit remaining correlations in the quantized feature space. In the context of our proposed CSI-INR scheme, we employ arithmetic coding given as:
$\bm{\bar{M}}^i_{out}=\mathcal{E}_e(\bm{\bar{M}^i}),\quad  \bm{\bar{M}}^i=\mathcal{D}_e(\bm{\bar{M}}^i_{out})$,
where $\mathcal{E}_e(\cdot)$ and $\mathcal{D}_e(\cdot)$ respresent the arithmetic encoder and decoder, respectively. To elaborate, we initially compute the distribution of quantized codes by assessing the frequency of each quantized modulation value within the training set. Subsequently, we utilize this distribution for arithmetic coding during test time. As shown in Table. \ref{exp_entropy}, entropy coding can provide an additional savings of up to $24.59\%$ in the number of transmitted bits without introducing any additional loss in performance.
\begin{table}[t]
\caption{Parameter settings for the DeepMIMO dataset}
\centering
\begin{tabular*}{0.85\linewidth}{ll}
\hline
\textbf{Parameters} & \textbf{Value}  \\
\hline
Operating frequencies& 3.5GHz \\
Bandwidth& 100MHz \\
Activated base station& BS 1 \\
Number of antennas & $N_t=32$ \\
Number of subcarriers & $N_c=32$ \\
Number of paths & $P=10$ \\
Users index & $R1 - R201$\\
\hline
\end{tabular*}
\label{dataset}
\end{table}

In summary, through the integration of quantization and entropy coding techniques, the processes executed within the transmitter of our INR-based CSI feedback schemes, as depicted in Eqn. \ref{sys_formulation}, can be articulated as follows: 
\begin{equation}
\begin{array}{ccc}
    \bm{M^i}=\mathcal{M}(\bm{H^i}), & \bm{\bar{M}}^i=\mathcal{Q}_q(\bm{M^i}), &\bm{\bar{M}}^i_{out}=\mathcal{E}_e(\bm{\bar{M}^i}),\\
\end{array}
\end{equation}
where the input CSI data undergoes modulation $\mathcal{M}(\cdot)$, quantization $\mathcal{Q}_q(\cdot)$ and entropy coding $\mathcal{E}_e(\cdot)$ to generate the codewords for low-cost CSI feedback. At the receiver, the symmetrical operations are sequentially executed as:
\begin{equation}
\begin{array}{ccc}
    \bm{\bar{M}}^i=\mathcal{D}_e(\bm{\bar{M}}^i_{out}), &                     \bm{\hat{M}}^i=\mathcal{D}_q(\bm{\bar{M}}^i),
&\bm{\hat{H}^i}={f}_{\bm{\theta}}(\bm{\hat{M}^i},\bm{x}),\\
\end{array}
\end{equation}
where the received codewords $\bm{\bar{M}_{out}}^i$ are subjected to a series of processing steps including entropy decoding $\mathcal{D}_e(\cdot)$, dequantization $\mathcal{D}_q(\cdot)$, and channel gain mapping $f_{\theta}(\cdot)$ to reconstruct the CSI data.

\subsection{CSI feedback strategy}
Finally, it is worth noting that we only feedback the modulations and assume both the transmitter and receiver are equipped with the shared base network, which serves as the codebook. In this way, only the modulated codewords undergo quantization and entropy coding, reducing the final communication cost. This aligns with the existing DL-based CSI-feedback solutions, wherein the transmitter and receiver have access to an autoencoder model, and the transmitted data only involves the quantized and entropy-coded bits of the codeword.

\subsubsection{Encoding at the user side} As shown in Algorithm. \ref{alg_feedback}, given a well-trained base network $\bm{\theta^*}$, user encodes different CSI samples into distinct modulated codewords within $s_{in}$ inner steps. More specifically, we firstly initialize a $\bm{M^i_0}$ for the $i$-th data point $\bm{d^i}=(\bm{x},\bm{H^i})$, and then perform $s_{in}$ times optimization steps as:
\begin{equation}
    \bm{M}_j^i\leftarrow\bm{M}^i_{j-1}-\alpha\nabla_{\bm{M^i_{j-1}}}\mathcal{L}_{mse}(\bm{\theta^*},\bm{M_{j-1}^i},\bm{d^i}),
    \label{inner_loop_inference}
\end{equation}
where the result $\bm{M_{s_{in}}^i}$ can then be quantized and entropy encoded for the final transmission. It is worth noting that the selection of $s_{in}$ is disentangled with the well-trained models, where $s_{in}$ can be different from $s$ for used in the inner-optimization training phase. It means users can dynamically modulate their CSI based on their own computation resources. Notably, empirical observations from our practical experiments indicate that even as few as $2-3$ gradient steps can yield significant improvements, while an increased number of steps tends to enhance the performance. Furthermore, a lightweight modulation network with only a single MLP layer can yield good performance. In light of these factors, alongside the adaptive optimization steps on the user side, the proposed CSI-INR scheme demonstrates efficiency and potential for an efficient feedback strategy.

\begin{table}[t]
\caption{Parameters setting for the model training}
\centering
\begin{tabular*}{0.85\linewidth}{ll}
\hline
\textbf{Parameters} & \textbf{Value}  \\
\hline
Batch size& $64$ \\
SIREN layer number & $L_t=10$\\
Hidden layer dimension & $d=512$\\
Sinusoidal parameter & $w_o=50$\\
Gaussian scale of FFN&$\sigma_b=10$\\
Inner optimization step for training& $s=3$ \\
Inner optimization step for inference& $s_{in}=3$ \\
Inner and outer learning rate & $\alpha=1e^{-2}$, $\beta=1e^{-6}$ \\
Optimizer & Adam optimizer\\
Loss function & MSE loss \\
\hline
\end{tabular*}
\label{training_setting}
\end{table}

\subsubsection{Decoding at the BS side}
As presented in Algorithm \ref{alg_de}, the received vector $\bm{\bar{M}^i_{out}}$ undergoes sequential processing through an arithmetic decoder and dequantizer to obtain $\bm{\hat{M}^i}$ for the subsequent decoding phase. 
Conversely, if quantization and entropy coding operations are absent, $\bm{{M}^i_{s_{in}}}$ is directly fed into the modulation and base networks for CSI reconstruction.

To be more specific, the dequantized $\bm{\hat{M}^i}$ or modulated codeword $\bm{{M}^i_{s_{in}}}$ is fed to the modulation network to map the modulated vectors $\bm{\phi^i}\triangleq\left[\bm{\phi^i_1},\cdots, \bm{\phi^i_{L_t}}\right]$ associated with each modulated SIREN layer of the base network $f_{\bm{\theta}}$, as Eqn. \ref{modulate}. This modulation results in the generation of distinctive CSI matrices, which can be formulated by:
\begin{equation}
    \bm{\hat{H}^i}=f_{\bm{\theta}}(\bm{M_{s_{in}}^i},\bm{x}).
\end{equation}

\section{Training and evaluation}
This section presents numerical experiments to evaluate the proposed CSI-INR method for the massive MIMO CSI feedback problem, assuming a high-frequency outdoor communication environment. In particular, we consider the Outdoor-1 Scenario of DeepMIMO datasets\cite{alkhateeb2019deepmimo} as the experimental environment, with an operating frequency of $3.5$ GHz. Models are trained until no further performance improvement is observed on a distinct validation set.

\subsection{Experimental setup}
The Outdoor-1 scenario in this study exemplifies a typical outdoor setting featuring two streets and a single intersection. Our dataset configuration, as detailed in Table. \ref{dataset}, involves the deployment of Base Station 1 (BS 1), equipped with $N_t=32$ isotropic antennas arranged in a ULA configuration. Specifically, we sample $40,000$ channel data points, which are partitioned into training, validation, and testing sets, with proportions of $70\%$, $10\%$, and $20\%$, respectively.

Training parameter configurations are summarized in Table \ref{training_setting}, where we use the Adam optimizer for back-propagation with a learning rate of $\beta=1e^{-6}$ in the outer loop, and a learning rate of $\alpha=1e^{-2}$ in the inner loop. We measure the reconstruction quality by the normalized mean square error (NMSE) over the space-frequency domain, defined as:
\begin{equation}
    \text{NMSE}\triangleq \mathbb{E}\left\{\frac{\|\bm{H}-\bm{\hat{H}}\|_2^2}{\|\bm{H}\|_2^2}\right\},
\end{equation}
where the expectation is taken across diverse CSI data.
\begin{figure}[t] 
    \centering
    \includegraphics[scale=0.445]{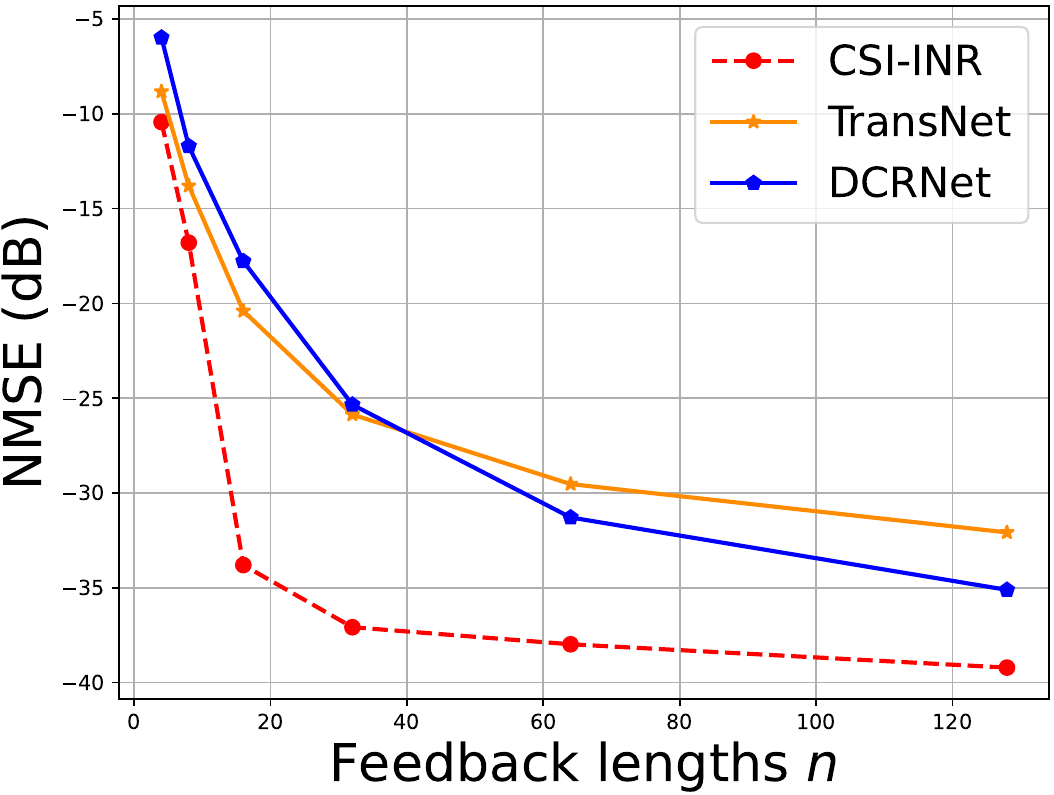}
    \caption{The NMSE performance of different schemes over varying feedback dimensions.}
    \label{general_performance}
\end{figure}

We employ DCRnet\cite{tang2022dilated}, TransNet\cite{cui2022transnet}, CSINet+\cite{guo2020convolutional} and DeepCMC\cite{mashhadi2020distributed} as the benchmarks. DCRnet\cite{tang2022dilated} and TransNet\cite{cui2022transnet} are chosen as benchmarks owing to their remarkable performance and advanced architectures based on dilated convolutional neural networks and transformers, respectively. CSINet+ scheme is one of the most representative feature learning-based CSI feedback schemes used as a benchmark in the literature. 
DeepCMC \cite{mashhadi2020distributed} is selected as the benchmark for its competitive performance and its utilization of entropy coding design, which can be used to compare with the proposed CSI-INR scheme when quantization and entropy coding are employed.

\subsection{General performance}
First, we validate the efficacy of viewing the CSI map as a neural function. Specifically, we compare the performance of the CSI-INR scheme with DCRnet\cite{tang2022dilated} and TransNet\cite{cui2022transnet} without quantization and entropy coding operations. The experimental performance of various schemes across diverse feedback dimensions is illustrated in Fig \ref{general_performance}. The figure reveals that the proposed CSI-INR yields a notable performance improvement of at least $1.61$ dB across all feedback dimensions. 
\begin{figure}[t] 
    \centering
    \includegraphics[scale=0.445]{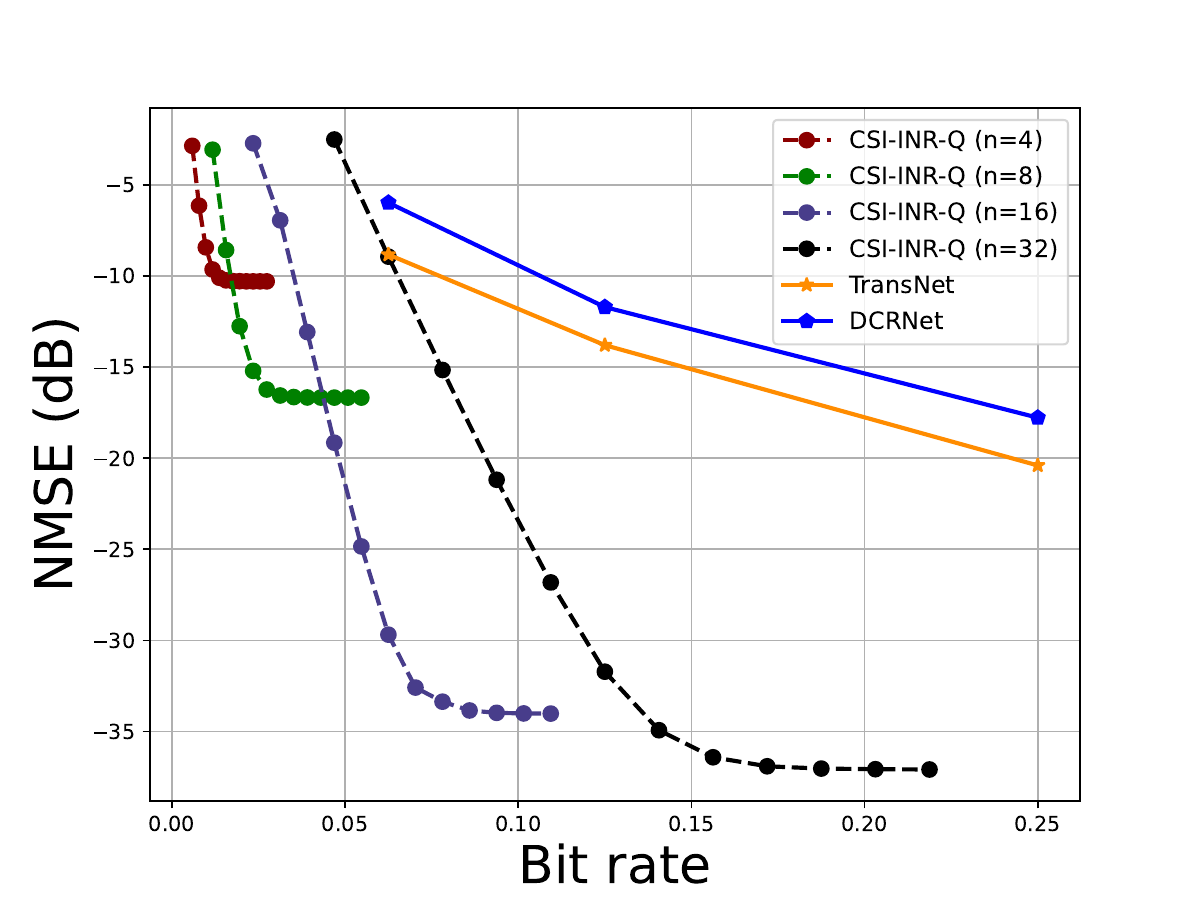}
    \caption{The rate-distortion performance of CSI-INR schemes across varying feedback lengths when the modulations are quantized with various bit widths. The data points on each curve denote the model performance across a range of quantization levels, ordered from highest ($14$ bits) to lowest ($3$ bits).}
    \label{exp_quant}
\end{figure}

Notably, the CSI-INR scheme demonstrates a pronounced performance advantage in low bandwidth scenarios, particularly when $n=16$ and $32$. In extremely limited bandwidth scenarios (such as $n=4$ or $n=8$), our approach exhibits a relatively smaller advantage. This can be attributed to the challenge posed by the modulation dimension being too low, which increases the difficulty of the INR model in accurately mapping each CSI data within a shared base network. In such cases, it is challenging for the INR model to distinguish between CSI data with a low-dimensional modulation vector for accuracy reconstruction. We can also observe that the performance gain of the CSI-INR scheme diminishes as the feedback length gradually extends. This phenomenon arises because, when the feedback length is sufficiently large, feature learning methods with powerful architectures such as CNN or transformer can attain satisfactory performance\cite{cao2021lightweight,guo2020convolutional,wen2018deep}. We further observe a pronounced decrease in the NMSE performance of the CSI-INR as the feedback length increases. In contrast, alternative methods exhibit a comparatively less significant reduction. This observation suggests that CSI-INR can better leverage the constrained feedback budget to encapsulate the essential information within the CSI dataset.

In summary, all the above findings demonstrate that considering the CSI map as a neural function attributes substantial performance benefits when compared with prior networks reliant on feature extraction for massive MIMO channel data.


\subsection{Effect of quantization}
We quantize the modulations of each CSI data from $3$ bits to $14$ bits for the proposed CSI-INR scheme, resulting in different bit rates for the CSI-INR model of each feedback dimension. The NSME performance over various bit rates is presented in Fig. \ref{exp_quant}, where the CSI-INR scheme with quantization operations is denoted as CSI-INR-Q. Generally, a reduced bitwidth of the CSI modulations results in more compressed codes, albeit at the expense of reduced NMSE. Compared with the benchmarks, where the TransNet and DCRNet are quantized with $32$ bits from their original design, we can observe that applying uniform quantization can improve the compression ratio by a factor of $\times 5$ for a target NMSE of $-15$dB. 

Significantly, the resultant bitwidths for CSI-INR models exhibit a trend of surprisingly low values. Specifically, for the CSI-INR scheme with $n=4$, satisfactory performance can be achieved with the quantization of $3$ bits, and for the CSI-INR scheme with $n=8$, $4$-bits quantization proves sufficient for acceptable results. Notably, a larger feedback dimension necessitates more bits for compression; for instance, $7$ bits and $8$ bits are required for $n=16$ and $n=32$ dimensions, respectively. This can be explained that the longer code length is attributed to the increased amount of information, necessitating a greater number of bits for quantization to achieve satisfactory performance.
\begin{figure}[t] 
    \centering
    \includegraphics[scale=0.445]{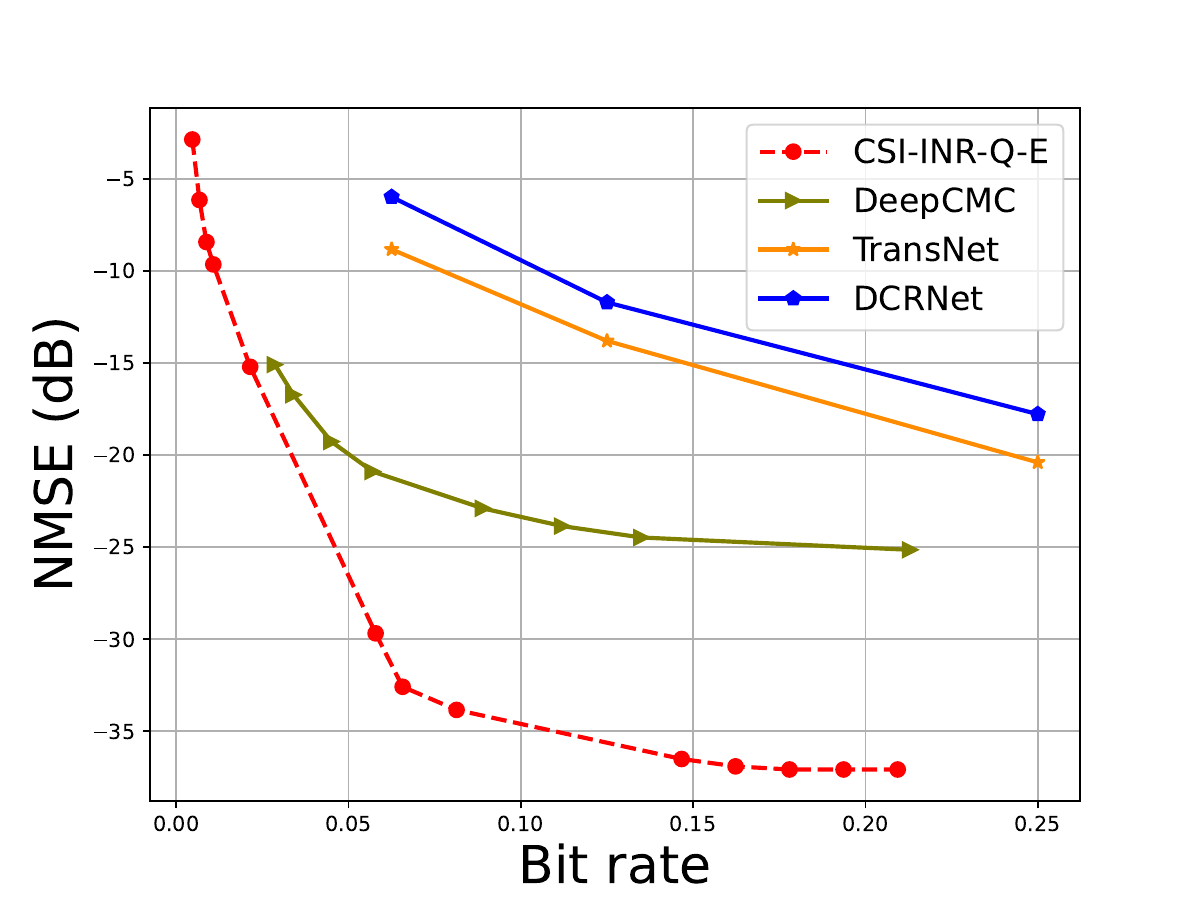}
    \caption{The comparisons of different schemes with quantization and entropy coding operations.}
    \label{exp_quant_entropy}
\end{figure}

It is essential to underscore that the quantization methodology utilized in the CSI-INR scheme operates independently of the training phase. This differs from previous approaches where the quantization is trained in an end-to-end fashion \cite{mashhadi2020distributed}, which necessitates the training of separate models tailored to various compression rates to achieve the desired rate-distortion trade-offs. Notably, the proposed CSI-INR scheme showcases robustness to different quantization methodologies, adapting effectively to different quantization levels once the CSI-INR model has been well-trained.

In summary, quantization within the CSI-INR scheme demonstrates significant superiority and effectiveness with substantial performance enhancements, despite employing a simple quantization method. The transmission requirement is minimal, involving only a few quantized bits, and its impact on feedback quality is negligible.

\begin{table}[t]
\caption{Bit rates and coding improvements from the entropy coding for different quantization levels of CSI-INR scheme.}
\begin{tabular}{|c|c|c|c|c|c|}
\hline
 \multicolumn{6}{|c|}{Models of various feedback lengths}\\
\hline
\multicolumn{2}{|c|}{Quantization levels}& $n=4$ & $n=8$ & $n=16$ & $n=32$\\
\hline
\multirow{2}{*}{$b=3$}&bit rates  & $0.00464$ & $0.00926$ & $0.01784$ & $0.03534$ \\
&coding gain & $20.82\%$& $20.97\%$ & $23.86\%$ & $24.59\%$ \\
 \hline
\multirow{2}{*}{$b=4$}&bit rates  & $0.006723$ & $0.01341$ & $0.02603$ & $0.05166$ \\
&coding gain  & $13.94\%$& $14.21\%$ & $16.69\%$ & $17.33\%$ \\
\hline
\multirow{2}{*}{$b=5$}&bit rates  & $0.00875$ & $0.017464$ & $0.03413$ & $0.06789$ \\
&coding gain  & $10.38\%$& $10.59\%$ & $12.63\%$ & $13.11\%$ \\
  \hline
\multirow{2}{*}{$b=6$}&bit rates  & $0.01074$ & $0.02145$ & $0.04209$ & $0.08383$ \\
&coding gain  & $8.31\%$& $8.47\%$ & $10.19\%$ & $10.58\%$ \\
 \hline
\multirow{2}{*}{$b=7$}&bit rates  & $0.01272$ & $0.02540$ & $0.04999$ & $0.09962$ \\
&coding gain  & $6.96\%$& $7.11\%$ & $8.59\%$ & $8.91\%$ \\
  \hline
\multirow{2}{*}{$b=8$}&bit rates  & $0.01468$ & $0.02933$ & $0.05784$ & $0.11533$ \\
&coding gain  & $6.01\%$& $6.14\%$ & $7.44\%$ & $7.73\%$ \\
 \hline
\end{tabular}
\label{exp_entropy}
\end{table}

\subsection{Effect of entropy coding}
In order to enhance the rate-distortion optimization, we incorporate entropy coding after quantization, where a pair of arithmetic encoder and decoder is employed to achieve lossless compression of the quantized CSI codewords, denoted as CSI-INR-Q-E.

For a fair comparison, we compare the performance of CSI-INR-Q-E with DeepCMC, which also incorporates quantization and entropy coding operations, alongside DCRNet and TransNet schemes. The experimental results are illustrated in Fig. \ref{exp_quant_entropy}, where the proposed CSI-INR-Q-E scheme provides a pronounced improvement of the CSI reconstruction quality (up to $11.23$ dB) across all bit rates. Meanwhile, for a target NMSE performance of $-25$ dB, CSI-INR-Q-E can attain this performance with a communication overhead reduced by a factor of $5$. This character holds particular significance, especially in light of the fact that feedback bits are commonly transmitted with a low coding rate and a small constellation size (such as QPSK) to ensure decoding accuracy. 

To elucidate the impact of entropy coding in CSI-INR schemes, we present a comprehensive analysis of bit rates and coding improvements resulting from entropy coding across a spectrum of quantization levels ranging from $b=3$ to $b=8$, as summarized in Table \ref{exp_entropy}. It is apparent that the utilization of entropy coding yields a notable enhancement in coding efficiency, achieving improvements of up to $24.59\%$ in terms of bit rate while maintaining the same reconstruction quality. The significant enhancement is particularly pronounced at lower quantization levels, correlating with increased feedback length. This implies that entropy coding exhibits superior performance within scenarios with longer feedback lengths and fewer quantization bit numbers.

In summary, incorporating entropy coding yields significant advantages for the CIS-INR method, with a notable enhancement in coding efficiency while preserving reconstruction quality. Compared to alternative approaches like DeepCMC, which integrates the quantization layer into the training process alongside the entropy encoder and advanced feature-learning architecture, CSI-INR achieves significant performance gains despite employing a relatively straightforward quantization and entropy coding scheme. Looking ahead, we expect to design advanced quantization and entropy coding methods to further bootstrap the performance.

\subsection{Performance over varying coding steps}
An additional advantage of the proposed CSI-INR scheme is its capability to dynamically select diverse modulation strategies based on individual computing resources. More specifically, users have the flexibility to opt for various inner steps to enhance the modulation optimization. The performance corresponding to different inner steps in the inference phase of the CSI-INR model is presented in Table \ref{exp_steps}.
\begin{table}[t]
\caption{NMSE performance (dB) of CSI-INR models over various feedback lengths and inference inner steps, where CSI-INR schemes are trained with $s=3$ and tested with various $s_{in}$.}
\centering
\begin{tabular}{|c|c|c|c|c|}
\hline
Models & $s_{in}=2$ & $s_{in}=3$ & $s_{in}=5$ & $s_{in}=8$\\
\hline
CSI-INR (n=32)  & $-24.41$& $-36.94$ & $-37.13$ & $-37.18$ \\
\hline
CSI-INR (n=64)  & $-34.18$& $-37.68$  &$-37.91$  &$-37.94$ \\
\hline
CSI-INR (n=128) & $-38.11$ & $-39.21$ & $-39.31$ &$-39.33$ \\
\hline
\end{tabular}
\label{exp_steps}
\end{table}

It is apparent that an increased number of inner optimization steps generally contributes to a better performance. Thanks to the MAML strategy, a reduced number of steps, such as $s_{in}=3$, is enough for the CSI-INR model to achieve satisfactory performance. Nevertheless, imposing a constraint on the step count to a value less than $2$ can result in significant performance degradation, particularly at smaller feedback dimensions, such as $n=32$. This may be due to the longer codelengths requiring fewer modulation steps. In summary, an increased code length requires fewer modulation steps to achieve a satisfactory performance.

It is essential to highlight that within this CSI-INR framework, only a single model undergoes training to accommodate diverse inner optimization step configurations. This adaptability allows users to dynamically select the number of steps with a single model based on their individual considerations, e.g., computational complexity or latency constraints. Concurrently, the provided model is capable of undergoing fine-tuning by employing a specific inner step configuration, thereby enhancing performance through the adjustment of specific inner optimization step configurations.
 \begin{figure}[t]
     \centering
    \subfloat[]{
    \label{loss}
     \begin{minipage}[t]{1\linewidth}
    \centering
    \includegraphics[scale=0.435]{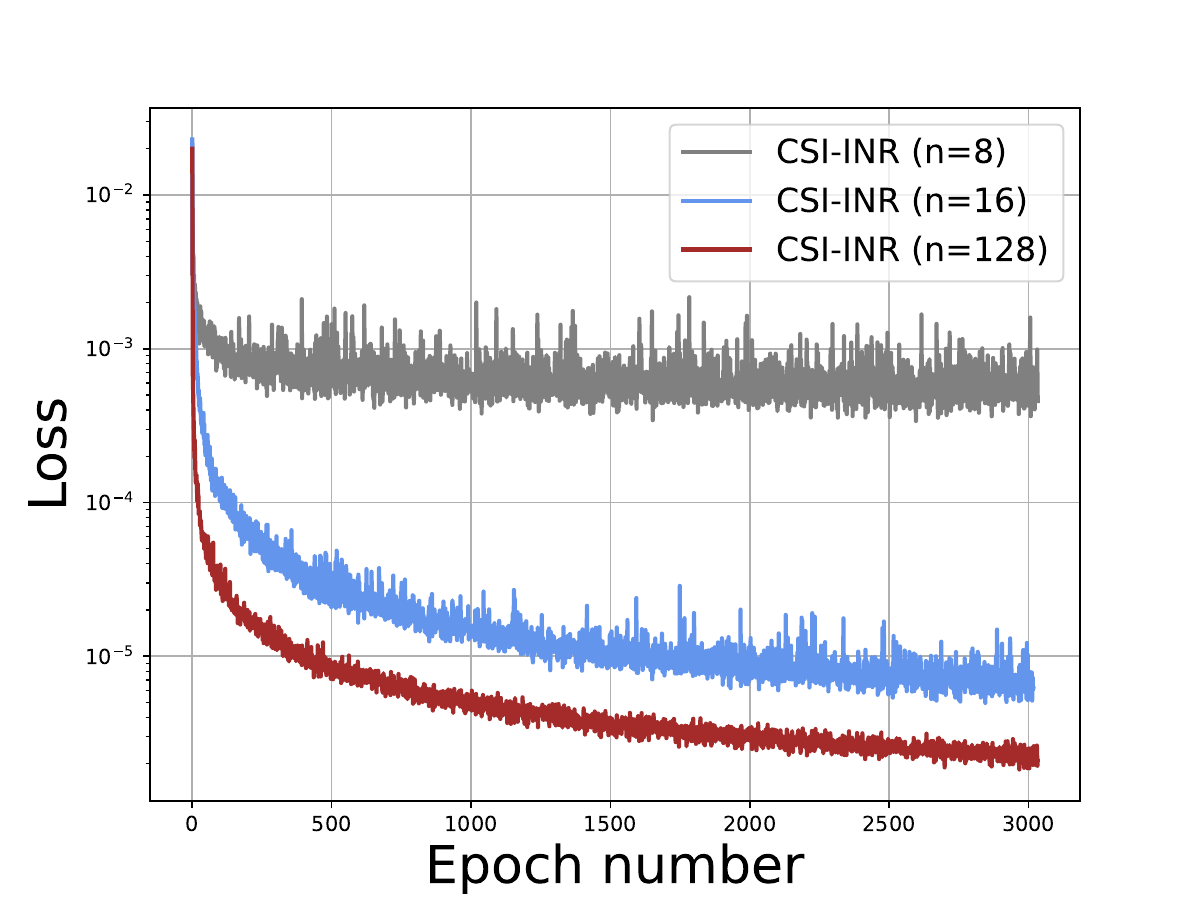}
     \end{minipage}%
     }%
     \hspace{-1mm}
     
     \subfloat[]{
     \label{nmse}
     \begin{minipage}[t]{1\linewidth}
     \centering
    \includegraphics[scale=0.435]{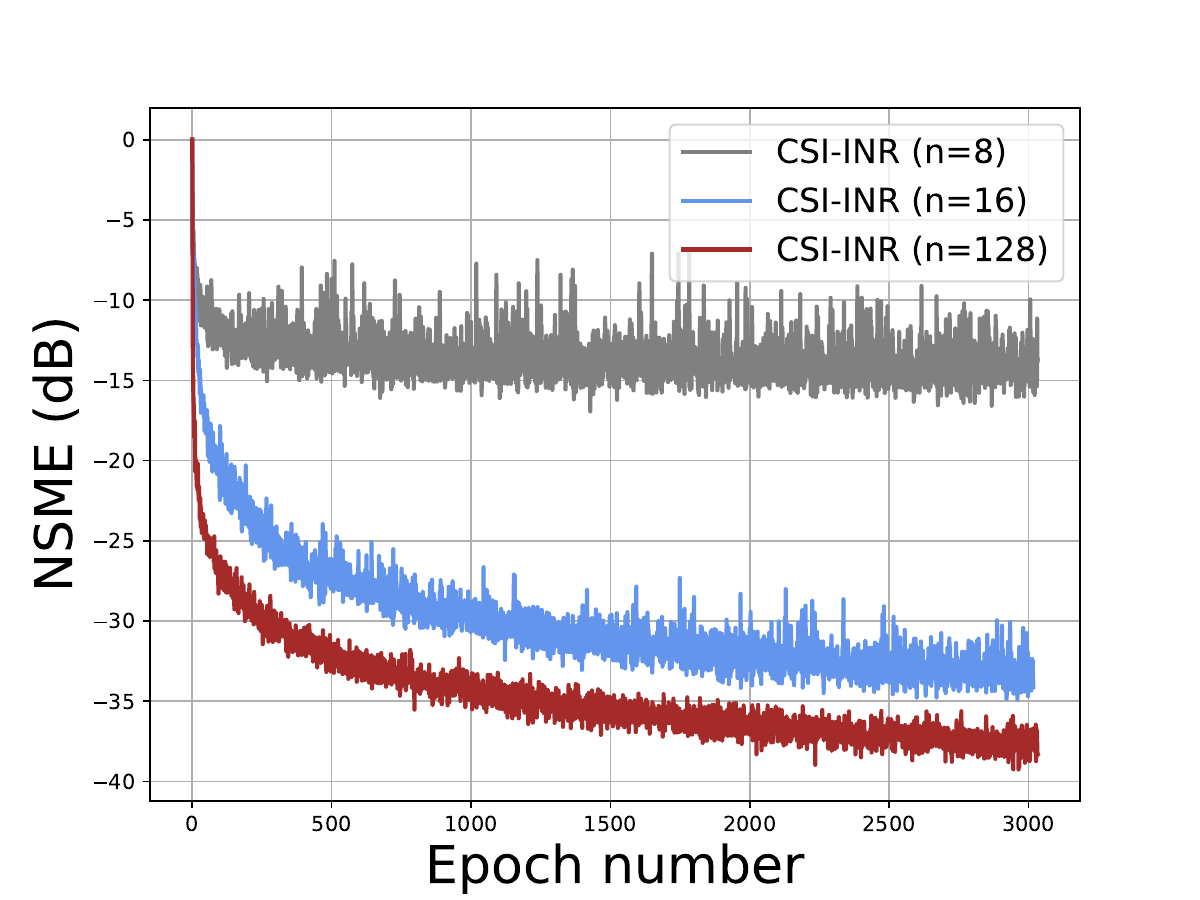}
     \end{minipage}%
     }%
     \caption{Loss function and NMSE trends of the proposed CSI-INR models during meta-learning process.}
     \label{convergence}
 \end{figure}
 
\subsection{Meta-learning Convergence}
Fast and stable convergence is another crucial aspect of the meta-learning process, where the computation of the second-order gradients within the MAML-based approach can yield instability for the training process. To elucidate the model convergence and training stability, we present the loss values and validation performance across the meta-learning process, as illustrated in Fig. \ref{convergence}.

We can observe that the proposed CSI-INR scheme can generally converge rapidly. Notably, the training process based on the MAML approach exhibits occasional instability, despite the model's tendency to recover from such instabilities rapidly. Some tricks, such as efficient first-order approximations for meta-learning, could mitigate these challenges. In summary, the MAML strategy exhibits consistent convergence and robust learning capabilities. This observation underscores the compatibility of fitting INR within a few steps, demonstrating the efficacy and efficiency of reconceptualizing CSI as neural functions.

\subsection{Ablation study over model efficiency}

\begin{table}[t]
\caption{Comparison of FLOPs, parameters and performance of different models during inference.}
    \begin{tabular}{ccccc}
    \toprule
              \multicolumn{5}{c}{\textbf{Feature learning-based schemes}}\\  
                      \bottomrule
   \multirow{3}{*}{CSINet+} & {FLOPs (M)}& \multicolumn{3}{c}{$44.58$}\\
& {Parameters (K)}& \multicolumn{3}{c}{$40.25$} \\
  & {NMSE (dB)}& \multicolumn{3}{c}{$-8.15$} \\ 
  \hline
\multirow{3}{*}{DeepCMC} & {FLOPs (M)}& \multicolumn{3}{c}{$556.91$}\\
& {Parameters (K)}& \multicolumn{3}{c}{$987.5$} \\
  & {NMSE (dB)}& \multicolumn{3}{c}{$-10.9$} \\ 
   \hline
\multirow{3}{*}{TransNet} & {FLOPs (M)}& \multicolumn{3}{c}{$37.85$}\\
& {Parameters (K)}& \multicolumn{3}{c}{$316.48$} \\
  & {NMSE (dB)}& \multicolumn{3}{c}{$-13.81$} \\ 
        \midrule
              \multicolumn{5}{c}{\textbf{INR-based schemes} ($s=3, n=8$)}\\  
        \midrule
         \multicolumn{2}{c}{\multirow{2}{*}{Hidden layer dimension }} &\multicolumn{3}{c}{Layer number}\\
         \cmidrule{3-5}
            & &  $L_t=8$ & $L_t=10$ & $L_t=12$\\
        \midrule
\multirow{3}{*}{$d=64$} & {FLOPs (M)}& $21.24$ & $29.62$ & $38.01$\\
& {Parameters (K)}& $27.47$ & $35.77$ & $44.09$ \\
  & {NMSE (dB)}& $-6.17$  &$-7.02$  &$-8.44$ \\
          \midrule  
\multirow{3}{*}{$d=128$} & {FLOPs (M)}& $84.42$ & $117.97$ & $151.53$\\
& {Parameters (K)}& $95.87$ & $128.89$ & $161.92$ \\
  & {NMSE (dB)}& $-9.68$  &$-10.23$  &$-12.74$ \\
  \midrule
\multirow{3}{*}{$d=256$} & {FLOPs (M)}& $336.62$ & $470.83$ & $605.05$\\
& {Parameters (K)}& $355.58$ & $487.17$ & $618.75$ \\
  & {NMSE (dB)}& $-12.13$  &$-15.96$  &$-17.26$ \\
    \bottomrule
    \end{tabular}
        \label{ab_arc}
\end{table}
In order to provide a comprehensive analysis of the trade-off between efficiency and performance concerning the proposed CSI-INR scheme, we conducted a detailed ablation study on the model architecture. Table. \ref{ab_arc} presents the model size, inference floating point operations (FLOPs), and the performance of the CSI-INR scheme across various configurations with parameters set to $s=3$ and $n=8$. 

As shown in the table, it is generally observed that increasing the number of layers and the dimension of hidden layers tends to enhance the performance of the proposed CSI-INR model. The parameter and FLOPs count of the CSI-INR model experiences a notable increase with the growth of the hidden layer dimension $d$, resulting in a more complex architecture associated with improved performance. 
In practical application, a suggested configuration is when $L_t=12$ and $d=256$, as it demonstrates satisfactory performance with manageable computational demands. 

Compared with other feature learning-based schemes, the proposed CSI-INR scheme demonstrates superior performance, albeit at the cost of increased computational resources. It is noteworthy that additional optimization opportunities exist regarding the complexity of our models, including exploring alternative hardware configurations, diverse implementations, and acceleration techniques tailored to enhance the efficiency of INR models. In summary, our proposed methodology underscores both practicality and efficiency, carefully considering computational complexity alongside performance metrics. 

\section{Conclusion}
We introduced an innovative INR-based CSI feedback methodology. This novel paradigm reconceptualizes the compression challenge associated with CSI data by leveraging implicit neural representations, which leverage the relation between the channel gains at different spatial coordinates, aligned with the physics governing the channel states. Specifically, we utilize neural functions to map the CSI coordinates to their respective channel gains. Instead of transmitting all parameters of neural functions, our proposed method adopts a MAML-based approach to modulate each CSI data into codewords for transmission. In addition, this meta-learning-based modulation approach exhibits remarkable and adaptive quantizability, where the proposed method can adapt to various quantization levels, showcasing its potential for robust performance even under low compression ratios. Another advantage of the proposed CSI-INR architecture lies in its potential for flexible modulation steps, where users can select diverse modulation steps within a single model based on their considerations for distinct target metrics, significantly increasing its practical applicability. The numerical results demonstrate that our proposed approach achieves state-of-the-art performance. 

Moving forward, several prospective directions for future exploration derived from this paper are outlined as follows:

\begin{itemize}
\item Performance optimization and data generalizability: Although the CSI-INR method exhibits competitive performance, there exists ample space for optimization in both performance and generalizability. Addressing the former concern over performance may entail the adoption of advanced quantization technologies, such as vector quantization or other entropy coding methods \cite{lightstone1997image}. Meanwhile, enhancing the MAML approach, specifically through implementing efficient first-order meta-learning strategies \cite{fallah2020convergence} can mitigate the latter issue of generalizability.

\item Inference efficiency: The proposed CSI-INR model adopts a lightweight architecture, utilizing only simple MLP layers. Nevertheless, the backpropagation operations in modulation operations introduce increased computational costs. A comprehensive exploration of the trade-off between computation and communication costs will be investigated in the future. Furthermore, advanced frameworks and modulation methods 
\cite{guo2023compression},in addition to the SIREN layer and MAML strategy, are anticipated to improve the framework's efficiency.

\item Analog feedback: An alternative approach is an analog CSI feedback scheme \cite{9053850}, which directly maps the CSI matrix to the input of a noisy feedback link.
Considering the flexibility and the efficiency of the INR-based model, we expect the formulation of an advanced analog feedback strategy that can achieve competitive performance over varying feedback link qualities.

\end{itemize}


\bibliographystyle{IEEEtran}
\bibliography{ref}

\vfill

\end{document}